# A Mechanism-Learning Deeply Coupled Model for Remote Sensing Retrieval of Global Land Surface Temperature

*Tian Xie[1†], Menghui Jiang[1†], Huanfeng Shen[1,2,3*], Huifang Li[1], Chao Zeng[1], Jun Ma[1], Guanhao Zhang[1], Liangpei Zhang[4*]*

[1] School of Resource and Environmental Sciences, Wuhan University, Wuhan 430079, China

[2] Key Laboratory of Geographic Information System of Ministry of Education, Wuhan 430079, China

[3] Key Laboratory of Digital Cartography and Land Information Application of the Ministry of Natural Resources, Wuhan 430079, China

[4] State Key Laboratory of Information Engineering in Surveying, Mapping and Remote Sensing, Wuhan University, Wuhan 430079, China

[†]These authors contributed equally: T. Xie, M. Jiang.
*Correspondence to: H. Shen –shenhf@whu.edu.cn, L. Zhang –zlp62@whu.edu.cn

**Abstract**:

Land surface temperature (LST) retrieval from remote sensing data is pivotal for analyzing climate processes and surface energy budgets. However, LST retrieval is an ill-posed inverse problem, which becomes particularly severe when only a single band is available. In this paper, we propose a deeply coupled framework integrating mechanistic modeling and machine learning to enhance the accuracy and generalizability of single-channel LST retrieval. Training samples are generated using a physically-based radiative transfer model and a global collection of 5810 atmospheric profiles. A physics-informed machine learning framework is proposed to systematically incorporate the first principles from classical physical inversion models into the learning workflow, with optimization constrained by radiative transfer equations. Global validation demonstrated a 30% reduction in root-mean-square error versus standalone methods. Under extreme humidity, the mean absolute error decreased from 4.87 K to 2.29 K (53% improvement). Continental-scale tests across five continents confirmed the superior generalizability of this model.

**Main text**

Land surface temperature (LST), as a pivotal parameter for Earth-atmosphere system interactions at regional and global scales, represents the most sensitive comprehensive indicator of spatiotemporal variations in surface energy and material fluxes. LST plays a crucial role in geophysical processes, including surface energy budget balance and global hydrological cycles[1,2], and is essential for identifying climate change trends and predicting potential extreme weather events[3,4,5,6,7]. With the continuous advancement of remote sensing retrieval and other technologies, LST has been extensively applied in climate monitoring, weather forecasting, agricultural management, ecological governance, urban planning, and water resource management[8,9]. Consequently, LST is vital for climate change mitigation and sustainable development research and has been designated as a priority measurement parameter by the International Geosphere-Biosphere Programme[10].

Satellite remote sensing is recognized as the most effective technique for LST retrieval, enabling spatially and temporally continuous LST measurements at global or regional scales. Thermal infrared bands, in particular, allow direct linkage between the radiative transfer equation (RTE) and LST, enabling the use of physics-based mechanistic models (MMs) for LST retrieval[11, 12, 13, 14, 15]. However, radiative transfer constitutes an extremely complex physical process influenced by atmospheric effects and other factors[9]. As the number of equations typically exceeds the number of unknowns in LST retrieval, this represents a classic ill-posed inverse problem. The challenge becomes particularly acute for single thermal infrared band configurations, where the limited observational information severely constrains high-accuracy LST retrieval. Researchers worldwide have developed various mechanism-based retrieval methods through approximate simplifications of the RTE[15, 16, 17, 18, 19, 20]. Nevertheless, these approaches can exhibit substantial errors under extreme conditions, such as high temperatures and high humidity. This mainly arises from the complexity, interactivity, and multiscale characteristics of the Earth system mechanisms and processes[21], which lead to certain assumptions and approximate processes that differ greatly from reality, making it difficult to accurately realize the precise modeling and accurate portrayal of the physical processes.

With the advancement of artificial intelligence technologies, machine learning (ML) has been increasingly applied in geophysical parameter retrieval[22, 23, 24, 25, 26, 27]. While ML methods excel at capturing nonlinear relationships through data-driven training iterations and prediction, the heavy reliance on training data and the limited interpretability severely restrict the generalizability under the case of data scarcity or distribution shifts. This limitation is exacerbated in LST retrieval because there are not enough actual surface temperature measurement sites. There are only a few dozen global ground observation sites, which are insufficient to support model building with high stability and generalization. Furthermore, the limited number of observation sites results in it being difficult to cover complex surface coverage scenarios Although regional-scale ML models linking in situ measurements with remote sensing observations have been attempted, they generally lack cross-regional transferability.

In typical applications, the mechanistic and data-driven models each possess distinct advantages and inherent limitations while exhibiting natural complementarity[28]. Coupling an MM with an ML model enables the synthesis of "rationalism" and "empiricism"[29], which is widely recognized as an inevitable development trend[21]. Scholars have proposed various coupled paradigms and frameworks for geophysical parameters[30, 31, 32, 33, 34], which can be primarily categorized into three basic approaches: mechanism-learning cascading, learning-embedded mechanism, and mechanism-integrated learning[21]. However, regarding LST retrieval, it is currently limited to a simple cascade approach, based on the use of a radiative transfer model to simulate training samples and then build a ML model[32, 35, 36, 37]. With the exponentially increasing data volumes, the challenges in detecting variability and interpreting extreme event patterns have intensified significantly. This primarily stems from insufficient utilization of information by both the mechanistic and data-driven models, resulting in underdeveloped complementary synergies[21]. Therefore, there is a pressing need to develop more tightly coupled modeling frameworks[38].

In this paper, we propose the novel mechanistic model-machine learning (MM-ML) deeply coupled framework for LST retrieval, integrating three fundamental coupling strategies to tightly interweave the physical mechanisms with an ML model. The framework begins with the generation of physically consistent training datasets through the simulation of atmospheric radiative transfer using 5810 atmospheric contours covering different climatic zones around the world. Meanwhile, we innovatively introduce a physically constrained multicomponent neural network (PCMCNN), which integrates a data-driven representation layer (DDRL), a physical parameter-guided layer (PPGL), and a physical process optimization layer (PPOL). Experimental validation shows that the framework can achieve a 30% reduction in root-mean-square error (RMSE) at the global scale and a 53% reduction in mean absolute error (MAE) under extreme humidity conditions, effectively overcoming the traditional mechanistic methods' limitations in complex environments. This breakthrough not only establishes a new paradigm for accurate inversion of surface temperature in complex environments but also provides reliable technical support for global climate change research and ecological environment dynamic monitoring.

## Results

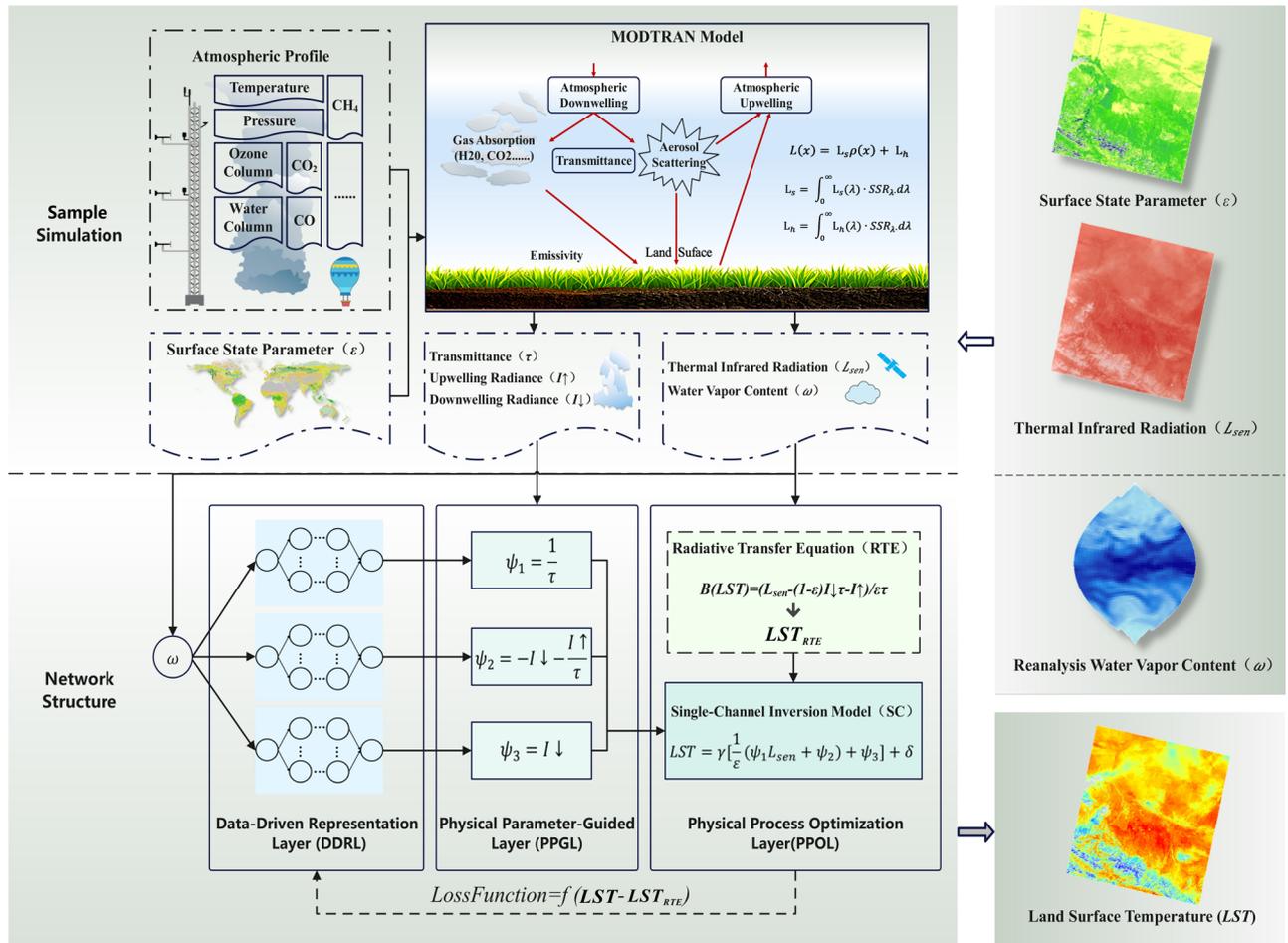

**Fig. 1 | Structure of the mechanistic model-machine learning coupling framework (MM-ML).** Schematic diagram of the coupling framework between the MM and ML. Inputting the atmospheric profile and surface state parameters, the MODerate resolution atmospheric TRANsmission (MODTRAN) model generates atmospheric parameters and radiative transfer parameters as inputs to the physically constrained multicomponent neural network (PCMCNN). The innovative proposed PCMCNN consists of a data-driven representation layer (DDRL), a physical parameter-guided layer (PPGL), and a physical process optimization layer (PPOL). The final well-trained MM-ML framework retrieves LST by inputting the surface state parameters, thermal infrared radiation, and atmospheric water vapor content.

## MM-ML coupling framework

To address the key challenges in global LST measurement, we propose the innovative MM-ML coupling framework that integrates multi-source information, including global reanalysis data, satellite observations, and meteorological station measurements. The framework incorporates a tightly coupled PCMCNN, which significantly enhances the accuracy and stability of LST retrieval through the advantages of dual coupling of the MM and ML.

The implementation scheme is illustrated in Fig. 1. Firstly, within the simulation sample generation, multi-source big data are synthesized, including radiosonde data, atmospheric profiles from meteorological reanalysis, and surface state parameters. The MODerate resolution atmospheric TRANsmission (MODTRAN) model is used to comprehensively simulate the key atmospheric processes in the radiative transfer process and generate high-precision simulation datasets to provide basic input data for the PCMCNN.

The PCMCNN adopts a progressively coupled architecture integrating physics and ML. In the DDRL, multiple ML layers are designed to mine feature representations in the data. The PPGL embeds the core atmospheric process models from single-channel (SC) algorithms, generating corresponding atmospheric functions based on the input water vapor content. Finally, the PPOL imposes consistency constraints through the RTE, aligning the mechanism-model-derived LST with the equation outputs. This multi-tiered design effectively incorporates physical knowledge, significantly enhancing the model's generalizability and physical interpretability.

The trained MM-ML framework takes thermal infrared radiance, surface parameters, and atmospheric water vapor content as inputs to retrieve the corresponding LST. Detailed implementation specifics are provided in the Methods section.

To comprehensively evaluate the proposed framework's accuracy and performance in LST

retrieval, we employ four metrics: mean absolute error (MAE), root-mean-square error (RMSE), bias, and the coefficient of determination ($R^2$). The multi-perspective validation confirms the model's reliability and superiority.

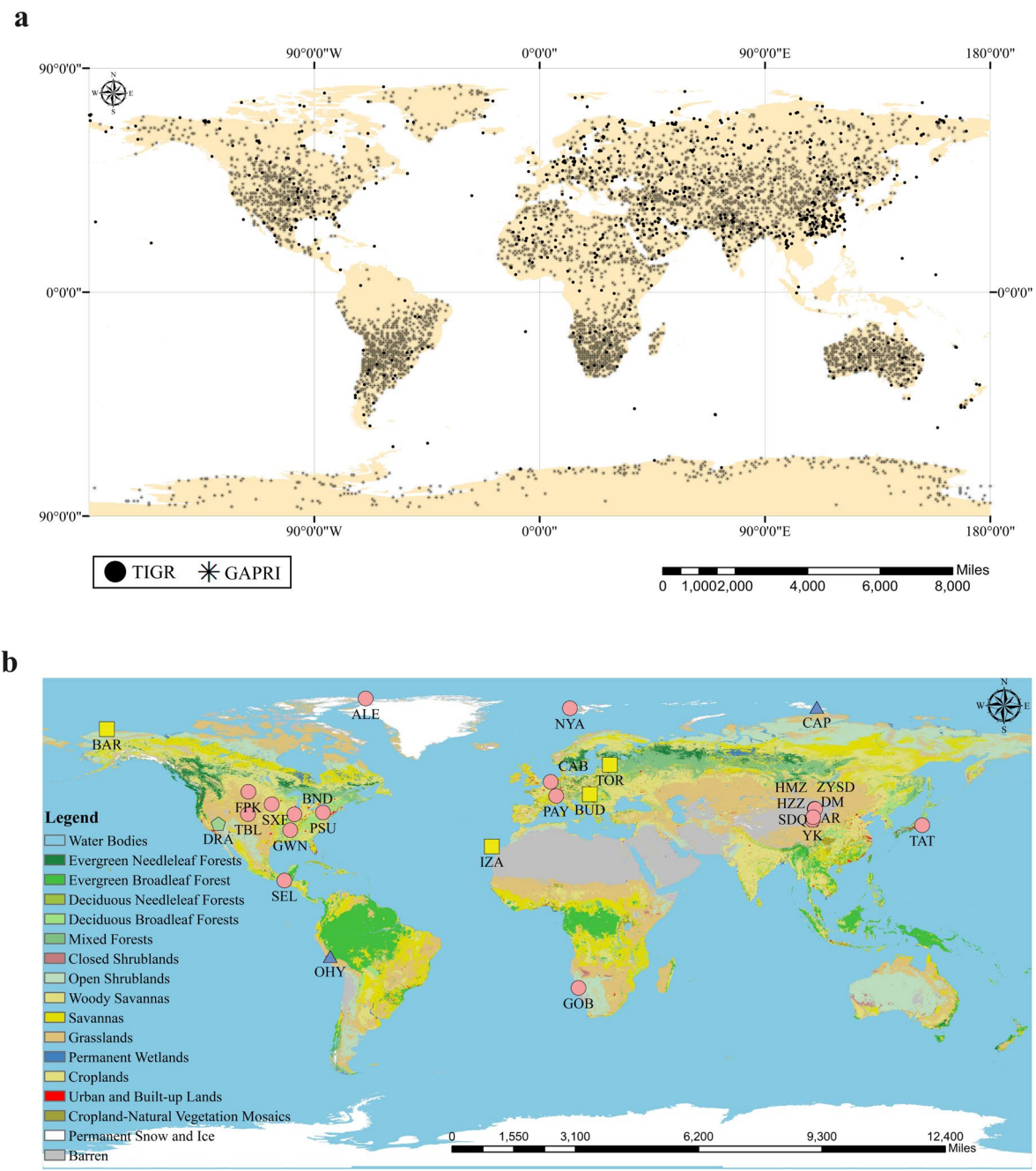

**Fig. 2 | Spatial distribution of the atmospheric profile datasets and the accuracy evaluation at 27 global ground sites across five continents. a**, The distribution of the TIGR and GAPRI atmospheric profile datasets over global land areas. **b,** Global land use/cover map for 2020 derived from the MCD12Q1 product (spatial resolution of 0.05°×0.05°), accompanied by the geographical locations of the ground sites. The study

region encompasses five continents, including global Baseline Surface Radiation Network (BSRN) sites, Surface Radiation Budget Network (SURFRAD) sites, and Heihe Integrated Observatory Network sites. The figure presents the mean absolute error (MAE) evaluation results for four models: the single-channel mechanism model (MM(SC)) (green), the radiative transfer mechanism model (MM(RT)) (yellow), the pure machine learning model (ML) (blue), and the MM-ML coupling model (red) at 27 global ground sites, highlighting the model with the highest accuracy. Detailed site information is available in Extended Data Table 1.

**Global site accuracy validation**

To systematically evaluate the model retrieval accuracy, we selected 27 representative ground sites across five continents (details in Extended Data Table 1) and compared the performance of four models: the single-channel mechanism model (MM(SC)), the radiative transfer mechanism model (MM(RT)), a pure ML model and the proposed MM-ML coupling model. As shown in Fig. 3a–d, these models exhibit significant performance disparities globally. Overall, MM-ML achieves a notably lower MAE, at 2.38 K, compared to MM(SC) (3.35 K), MM(RT) (3.20 K), and pure ML (3.05 K). The details reveal that MM-ML leverages ML's nonlinear fitting capability to significantly improve the prediction accuracy in the median temperature ranges while demonstrating superior robustness in the extreme temperature zones. In contrast, MM(SC) shows sparser scatterplot distributions with higher dispersion around the diagonal, indicating instability and precision limitations. MM(RT) exhibits systematic overestimation in the high temperature regions, which degrades the overall accuracy. Although the pure ML model performs well in the median temperature regions, its precision markedly declines under extreme temperatures, thereby compromising reliability.

The spatial distribution of the ground sites is shown in Fig. 2b, with the MAE results across the different sites shown in Fig. 3e. Of the 27 sites, the MM-ML model is superior at 19 sites and is mainly in second place at the remaining sites; the MM(RT) is superior at four sites; and the MM(SC) and pure ML model show a superior accuracy at only two sites each. Exceptions can be observed at specific sites, such as DRA, where both the MM-ML and ML models underestimate LST, potentially due to local surface cover characteristics or unique vegetation phenology[39]. However, overestimation by the MM(SC) and MM(RT) at this site partially compensates for the errors, contributing to the accuracy metrics.

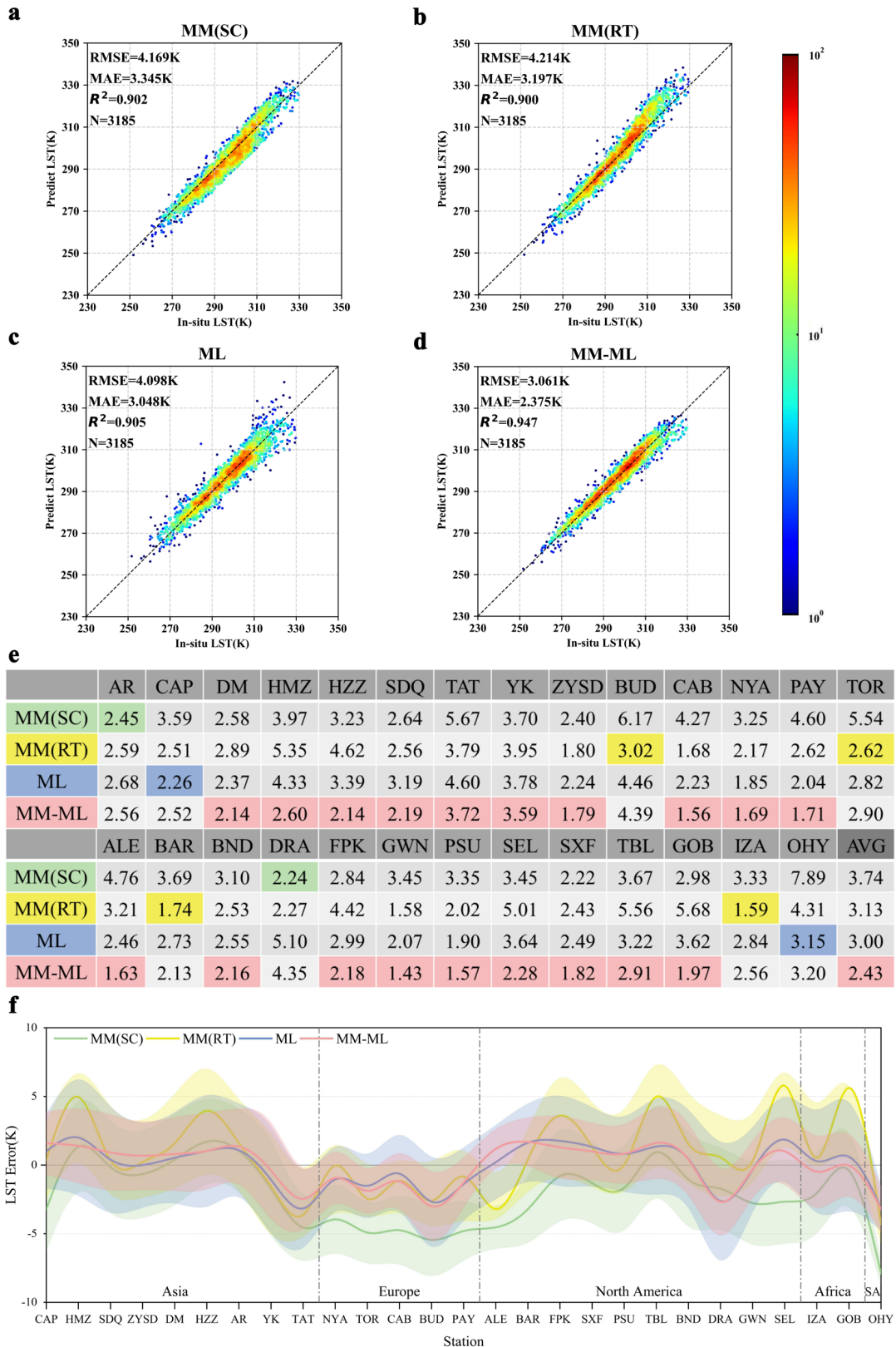

**Fig. 3 | Performance of the different models on the global test dataset of 27 sites. a,** Validation accuracy of MM(SC). **b,** Validation accuracy of MM(RT). **c,** Validation accuracy of pure ML. **d,** Validation accuracy of MM-ML. **e,** Accuracy evaluation results of the four models at the 27 global ground sites, showing the MAE for each model. **f,** Bias statistics of the four models at sites across Asia, Europe, North America, Africa, and South America (indicated in the figure by SA), with the sites sorted in decreasing order of latitude.

To investigate the regional and latitudinal impacts on model performance, we statistically analyzed the bias distribution across the continental sites (Fig. 3f). The results show systematic overestimation tendencies in Asia, North America, and Africa, contrasted with underestimation in Europe and South America. MM-ML maintains a lower bias in most cases, demonstrating strong global adaptability. The physics-based models (MM(SC)/MM(RT)) exhibit a geographically dependent performance: MM(SC) performs best at mid-latitude Northern Hemisphere sites (36–38°, e.g., SDQ and YK), but shows large deviations in the high- and low-latitude regions, which may stem from the insufficient data coverage in the high- and low-latitude areas. MM(RT) outperforms its inland counterparts at the coastal sites, potentially benefiting from stable marine atmospheric conditions. In contrast, MM-ML retains a superior/second-place accuracy under complex conditions through the physical constraints, substantially reducing the training data dependency while enhancing global generalization. The introduction of the PPGL and PPOL not only enhances the adaptability of the model to the complex radiometric characteristics of the different regions but also effectively compensates for the bias in the case of data scarcity or insufficient sampling. This coupled approach demonstrates strong potential for high-precision LST retrieval across diverse geographical regions, offering an effective solution for global-scale applications.

**Extreme condition validation**

LST retrieval accuracy significantly degrades under extreme climatic conditions (e.g., high temperatures, high humidity), posing a critical challenge for the traditional MM and pure

ML models. MMs struggle to capture complex environmental variations, due to oversimplified assumptions, while ML models exhibit instability in extreme temperature zones, due to the heavy data dependency. We systematically evaluated the four models to analyze their adaptability and precision under extreme atmospheric water vapor and LST conditions. Validation on the top and bottom 10% water vapor and LST samples (Fig. 4a, b) demonstrates the MM-ML model's superior stability, with errors of 2.35 ± 1.85 K and 2.29 ± 1.82 K (low/high water vapor), and 2.08 ± 1.69 K and 3.92 ± 2.69 K (low/high LST). In contrast, the physics-based models of MM(SC) and MM(RT) show nearly double the error under high water vapor (4.87 ± 2.81 K) and high LST (5.20 ± 3.51 K) conditions, exposing the limitations from the oversimplified physical parameterizations. The pure ML model's sensitivity to simulated data yields errors of 3.18 ± 2.81 K (low LST) and 5.23 ± 4.03 K (high LST), and the errors under high temperature conditions are close to twice those of MM-ML, highlighting the limitations of the pure ML model under extreme climatic conditions.

Visual comparisons of the retrieval results under high water vapor and LST conditions are shown in Fig. 4c, d. Meanwhile, Extended Data Figs. 2–3 provide a visualization comparison for the low water vapor and LST conditions. Further comparisons are made with the results of the Moderate Resolution Imaging Spectroradiometer (MODIS) LST product, which is widely recognized as having a superior accuracy, despite the lower spatial resolution of the MODIS LST product (MOD11A1 spatial resolution of 1 km). The results show that the MM-ML results are highly consistent with the MODIS LST product and show some advantages under all four extreme conditions. In contrast, the MM(SC) and MM(RT) models exhibit significantly amplified errors in the high temperature and high humidity scenarios, while the pure ML model displays abrupt regional discontinuities, due to its dependency on training data, further illustrating its inadequacy in complex conditions.

In summary, the MM-ML model, through deep coupling of the mechanistic and data-driven methods, effectively overcomes the limitations posed by insufficient simulation data under extreme atmospheric water vapor and surface temperature conditions. MM-ML maintains a high accuracy across various extreme conditions, significantly outperforming the pure traditional mechanistic and ML models, and shows excellent adaptability and robustness in complex environments. As such, MM-ML not only provides an efficient and precise solution for LST retrieval but also lays a solid foundation for further advancements in LST retrieval under extreme climatic conditions.

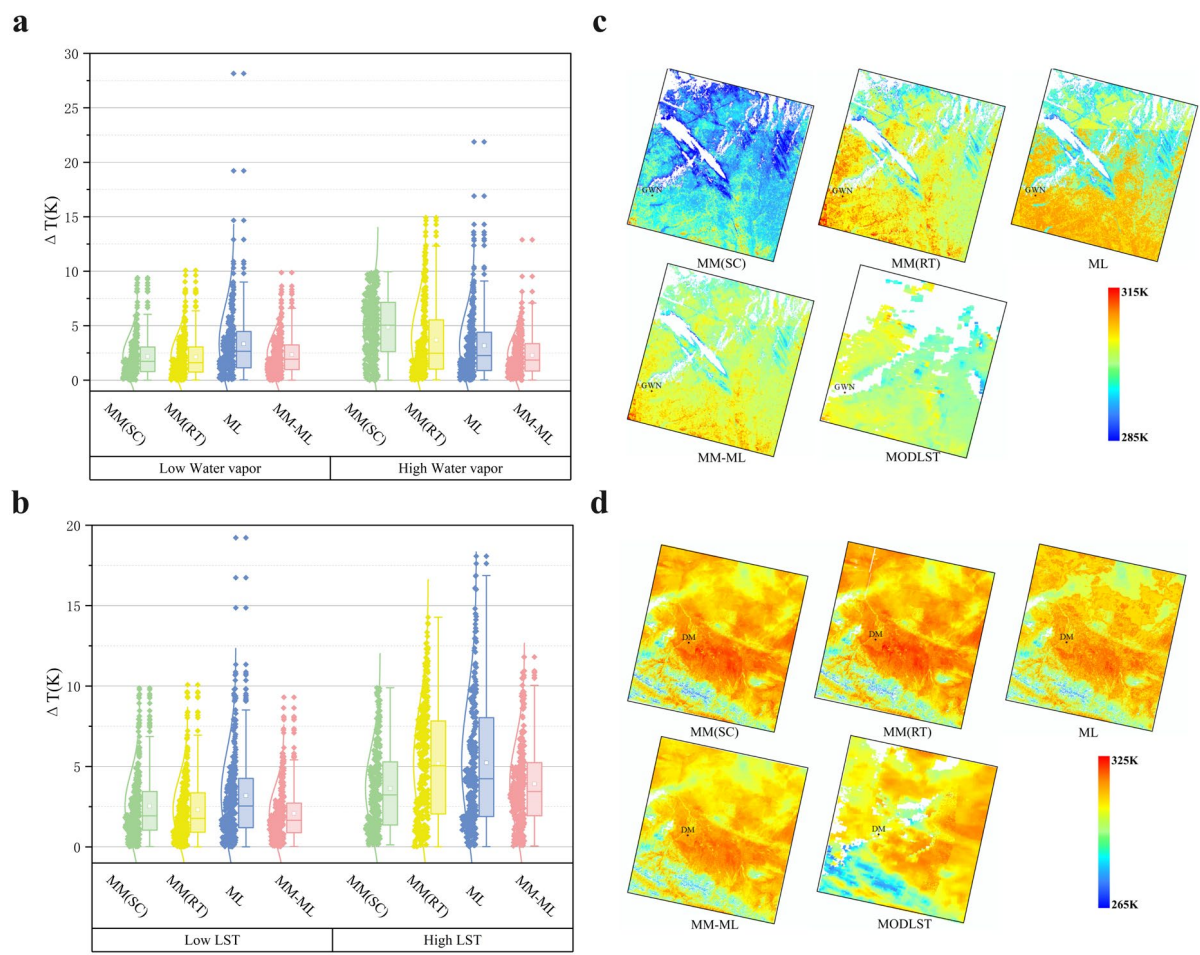

**Fig. 4| Validation of the MM(SC), MM(RT), pure ML, and MM-ML models under extreme conditions. a,** Box plots of the absolute error statistics for the top 10% of sites with higher atmospheric water vapor content (high water vapor) and the bottom 10% with lower atmospheric water vapor content (low water vapor) in the ground site dataset. **b,** Error box plots for the top 10% of sites with higher land surface temperatures (high LST) and the bottom 10% with lower land surface temperatures (low LST) in the ground site dataset. **c,** Validation results of the four models against the MODIS LST product (MODLST) at the GWN site in North America on September 19, 2015, under high water vapor content (4.093 g/cm²). **d,** Validation results of the four models against MODLST at the DM site in Asia on April 17, 2019, under high LST (312.506 K).

**Model sensitivity validation**

Atmospheric water vapor content, at-sensor radiance, and surface emissivity are critical input variables for LST retrieval models, with their variations directly impacting the model's accuracy and robustness. In this study, to further explore the sensitivity of these input parameters, we systematically evaluated the performance of the MM(SC) and MM-ML models under different input perturbation conditions using simulated datasets. Notably, the MM(RT) relies on Landsat LST products for retrieval, and the pure ML model depends on ground site data, making them unsuitable for sensitivity analysis with simulated datasets.

By quantifying the LST variations before and after the controlled input perturbations, the statistical results (Fig. 5a, b) reveal that a 5% variation in atmospheric water vapor induces statistically significant deviations in MM(SC) output: 0.08 ± 0.37 K (positive bias) and −0.07 ± 0.35 K (negative bias), with corresponding RMSE values of 0.24 K and 0.23 K. In contrast, MM-ML demonstrates enhanced noise resistance, showing smaller deviations (0.13 ± 0.31 K and –0.10 ± 0.26 K) and lower RMSE (0.18 K and 0.16 K). Under ±5% sensor-reaching radiance errors, MM(SC) exhibits a larger bias (4.90 ± 1.60 K and –5.08 ± 1.68 K), while MM-ML reduces these errors to 3.91 ± 1.02 K and –4.03 ± 1.07 K, confirming its superior radiance error robustness. For the emissivity perturbations, MM-ML outperforms MM(SC), with bias values of –1.87 ± 0.69 K and 2.06 ± 0.77 K versus –2.28 ± 0.59 K and 2.52 ± 0.65 K, respectively.

The proposed MM-ML model consistently demonstrates higher stability and robustness, compared to the traditional MM(SC) model, across all the tested input perturbations (water vapor, radiance, emissivity), and particularly maintaining lower errors under the case of large disturbances. This highlights its exceptional noise immunity and precision in complex environments, positioning it as a promising solution for remote sensing applications demanding high accuracy and operational stability.

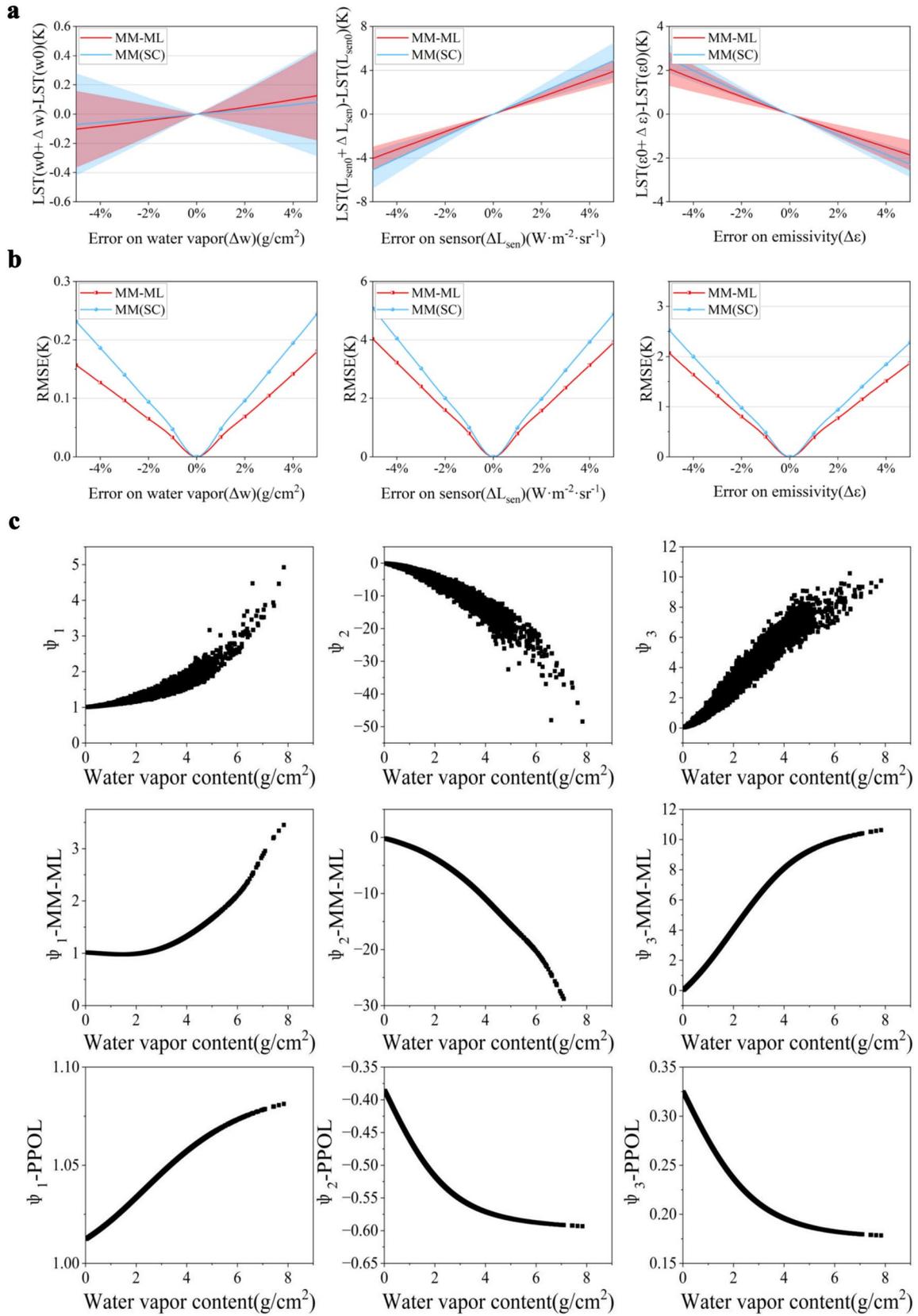

**Fig. 5| Sensitivity analysis for the MM-ML and MM(SC) parameters, and the ablation experiment results of the coupled model. a,** The impact of atmospheric water vapor content, satellite at-sensor radiance, and land surface emissivity errors on the LST retrieval errors. **b,** The impact of atmospheric water vapor

content, satellite at-sensor radiance, and land surface emissivity errors on the RMSE of the LST retrieval. **c,** Correlation plots of the three atmospheric function parameters ($\psi_1$, $\psi_2$, $\psi_3$) predicted by the MM-ML model and the single PPOL, with atmospheric water vapor content.

**Ablation experiments**

To investigate the interactions between the hierarchical components of the PCMCNN, we conducted ablation experiments assessing the individual components, i.e., the PPGL and the PPOL. The results show that when only the PPGL is activated, the model primarily replaces the empirical coefficient estimation for atmospheric functions ($\psi_1$, $\psi_2$, $\psi_3$) in the traditional physics-based approaches. While this improves the atmospheric function accuracy (see Extended Data Fig. 4), the accuracy of the final LST retrieval is better than that of the MM(SC) but does not exceed that of the full MM-ML model (Extended Data Table 2), because of the failure to integrate the constraints on the physical processes that follow.

In the PPOL-only configuration, where atmospheric function labels were not used, the loss functions were directly constructed using LST derived from the RTE. The experimental results (Fig. 5c) reveal that while this setup partially captures the physical correlations between water vapor and $\psi_1/\psi_2$, it exhibits significant deviations in $\psi_3$, compared to the established physical models, under complex atmospheric conditions.

The ablation experiments demonstrate that neither the PPGL nor PPOL alone can fully replicate the performance of the integrated MM-ML framework. While the PPGL enhances atmospheric function estimation through physics-guided learning, its lack of system-level constraints limits the final LST accuracy improvement. Conversely, the PPOL's direct optimization of the RTE struggles to maintain physical consistency in complex atmospheric scenarios without explicit mechanism guidance. The MM-ML framework's superiority stems from its synergistic integration of both components: the PPGL establishes physically meaningful atmospheric function relationships through mechanism-embedded learning, while the PPOL enforces system-level radiative balance constraints through physics-optimized loss functions. This dual-physics integration enables MM-ML to simultaneously overcome the limitations of the pure mechanistic and ML models, achieving unprecedented accuracy in LST retrieval across diverse environmental conditions. The results confirm that hierarchical physics integration—from localized atmospheric processes to global energy balance constraints—is critical for developing robust, interpretable Earth observation models that combine mechanistic understanding with data-driven flexibility.

**Discussion**

LST retrieval has long faced the dual challenges of data scarcity and complex environmental variability. In this paper, we have proposed the MM-ML model, which is a novel framework that deeply couples physical mechanisms with ML. Compared with the traditional approaches, MM-ML reduced the LST retrieval RMSE by 1.108 K over MM(SC), 1.153 K over MM(RT), and 1.037 K over the pure ML model, demonstrating its precision superiority.

The framework's innovation lies in its hierarchical physics integration. The PPGL alleviates the reliance on massive training data by embedding physical relationships into the learning processes, enabling robust performance across diverse geographies and extreme climatic conditions. Simultaneously, the PPOL makes the model close to the actual physical scene, effectively preventing the ML model from falling into locally optimal solutions. This dual physical constraint not only avoids the common overfitting problem of ML models but also enhances the robustness of the perturbation of the input parameters.

By harmonizing mechanism rigor with ML flexibility, the proposed MM-ML framework transcends the limitations of the existing methods in extreme conditions (e.g., high humidity), providing an accurate, adaptable tool for global LST monitoring. This paradigm shift in LST retrieval technology offers critical support for climate change research and environmental governance, bridging the gap between physical theory and data-driven innovation.

**Methods**

**Remote sensing data and input parameters.** In this study, Band 10 (10.6-11.2 μm) of the Landsat 8 Thermal Infrared Sensor (TIRS) was prioritized for LST retrieval based on its demonstrated superiority in signal-to-noise ratio characteristics. While band 11 (11.5–12.5 μm) was excluded owing to stray light contamination and calibration uncertainties. Land surface emissivity was obtained from the ASTER GEDv3 product, where normalized difference vegetation index (NDVI) thresholding was applied to estimate the emissivity for diverse land covers (vegetation, bare soil, etc.). Atmospheric water vapor, which is a critical input parameter, was derived from the National Centers for Environmental Prediction (NCEP) reanalysis dataset. This dataset integrates ground observations, ship data, radiosonde data, and satellite retrievals with a 6-hour temporal resolution. Linear interpolation ensured temporal alignment with the Landsat 8 overpass times.

**Atmospheric profiles.** Two atmospheric profile databases—Global Atmospheric Profiles derived from Reanalysis Information (GAPRI) and Thermodynamic Initial Guess Retrieval (TIGR)—were used to simulate the radiative transfer process. The GAPRI database provides 8324 global vertical atmospheric profiles, with a spatial resolution of 0.75° × 0.75° and a temporal resolution of 6 hours. It includes profiles from different climatic zones (e.g., tropical, mid-latitude, and subarctic regions) and varying levels of atmospheric precipitable water[40]. The TIGR database comprises 2311 representative profiles extracted from 80,000 radiosonde observations, encompassing diverse air masses (tropical, mid-latitude summer/winter, subarctic summer/winter)[41, 42]. Only land-based profiles were selected, to exclude oceanic influences (the spatial distributions are shown in Fig. 2a).

**Validation dataset.** The validation dataset was derived from three networks: the Baseline Surface Radiation Network (BSRN), the Surface Radiation Budget Network (SURFRAD), and the Heihe Integrated Observatory Network in the Heihe River Basin. The BSRN was established by the World Radiation Monitoring Center (WRMC) to support the World Climate Research Program (WCRP) and other scientific initiatives. It provides high-quality observations of shortwave and longwave surface radiation fluxes at high sampling rates, with a focus on monitoring the critical changes in surface radiation balance, and is widely used in global climate studies[43]. The SURFRAD network, which is operated by the National Oceanic and Atmospheric Administration (NOAA), covers multiple climate regions across the United States and provides continuous observational data, including upward and downward thermal infrared radiation[44]. Seven sites, located in Montana, Colorado, and other states, were selected for the validation. The Heihe Integrated Observatory Network, located in the arid inland of northwestern China, features diverse land surface types, including agricultural fields, deserts, wetlands, and alpine grasslands[45, 46, 47]. Given the inconsistent regional coverage of some sites, we selected surface temperature observation data from seven automatic meteorological stations within the Heihe Integrated Observatory Network for the model validation. For this study, sites from the major climatic zones were selected as part of the global validation dataset. The BSRN, SURFRAD, and Heihe Integrated Observatory Network ground sites were categorized into five regions based on geographical location: Asia, Europe, North America, Africa, and South America. The distribution of the ground sites is shown in Fig. 1b. LST was estimated using the following equation:

$$LST = \left[\frac{I\uparrow - (1-\varepsilon_b)\cdot I\downarrow}{\varepsilon_b \cdot \sigma}\right]^{1/4} \quad (1)$$

Where $I\uparrow$ and $I\downarrow$ represent the upward and downward thermal infrared irradiance, respectively; $\sigma$ is the Stefan-Boltzmann constant ($\sigma = 5.6705 \times 10^{-8}\ W\cdot m^{-2}\cdot K^{-4}$); and $\varepsilon_b$ is the broadband emissivity derived from 8-day MODIS LST emissivity products[48]. The emissivity data were matched to the validation sites using the temporally closest MODIS acquisitions.

**Fundamentals of LST retrieval.** According to Planck's radiation law, all objects with temperatures above absolute zero emit electromagnetic radiation. Under local thermodynamic equilibrium conditions, the spectral radiance of a blackbody at wavelength $\lambda$ and temperature $T$ can be expressed as:

$$B_\lambda(T) = \frac{c_1}{\lambda^5 \left[\exp\left(\frac{c_2}{\lambda T}\right) - 1\right]} \tag{2}$$

where $c_1 = 1.191 \times 10^8$ W μm$^4$sr$^{-1}$m$^{-2}$ and $c_2 = 1.439 \times 10^4$ μm $K$, which represents the radiation constants. For natural surfaces that are not ideal blackbodies, the specific emissivity is defined as the ratio of the radiant luminance of the object to the radiant luminance of the blackbody at the same temperature. When atmospheric effects are neglected, LST can be directly retrieved through $\varepsilon$ and $B_\lambda(T)$.

**MM(SC).** The MM(SC) utilizes a first-order Taylor series expansion of the Planck function around a specific temperature value and computes the LST using the following equation[12]:

$$LST = \gamma \left[\frac{1}{\varepsilon}(\psi_1 L_{sen} + \psi_2) + \psi_3\right] + \delta \tag{3}$$

Where $L_{sen}$ denotes the sensor-reaching radiance. Parameters $\gamma$ and $\delta$ are derived from Planck function parameterization:

$$\gamma = \left\{\frac{c_2 L_{sen}}{T_{sen}^2}\left[\frac{\lambda_{eff}^4}{c_1}L_{sen} + \lambda_{eff}^{-1}\right]\right\}^{-1} \tag{4}$$

$$\delta = -\gamma L_{sen} + T_{sen} \tag{5}$$

Where $T_{sen}$ represents the at-sensor radiance brightness temperature; $\lambda_{eff}$ is the effective wavelength; and $\psi_1$, $\psi_2$, and $\psi_3$ are the three atmospheric function parameters, defined as follows:

$$\psi_1 = \frac{1}{\tau}; \ \psi_2 = -I\downarrow - \frac{I\uparrow}{\tau}; \ \psi_3 = I\downarrow \qquad (6)$$

Where $\tau$ denotes the atmospheric transmittance; and $I\uparrow$ and $I\downarrow$ represent the upwelling and downwelling atmospheric radiance, respectively. The practical model in the SC algorithm approximates the atmospheric function parameters through quadratic polynomial fitting with the atmospheric water vapor content $\omega$, which can be expressed in matrix form as:

$$\begin{bmatrix}\psi_1\\\psi_2\\\psi_3\end{bmatrix} = \begin{bmatrix}c_{11} & c_{12} & c_{13}\\c_{21} & c_{22} & c_{23}\\c_{31} & c_{32} & c_{33}\end{bmatrix}\begin{bmatrix}\omega^2\\\omega\\1\end{bmatrix} \qquad (7)$$

In this study, we recalibrated the universal atmospheric function model using updated atmospheric profile datasets, resulting in optimized matrix coefficients.

**MM(RT).** When satellite-borne infrared sensors observe land surfaces along the line of sight, the sensor-reaching radiance comprises both surface-emitted radiation and atmospheric contributions. Under cloud-free conditions, the top-of-atmosphere (TOA) radiance can be expressed through the RTE:

$$L_{sen} = \varepsilon B(T_{sen})\tau + (1-\varepsilon)I\downarrow\tau + I\uparrow \qquad (8)$$

Where $B(T_{sen})$ denotes the blackbody radiance at LST $T_{sen}$. The Landsat Collection 2 Level-2 LST product, developed by the United States Geological Survey (USGS), which is based on the RTE, integrates TOA brightness temperature, TOA reflectance, ASTER Global Emissivity Dataset (GED) data, ASTER NDVI data, and atmospheric profiles from reanalysis datasets to produce high-precision standardized LST retrievals[49]. This product currently demonstrates a comparatively high spatial resolution among existing LST datasets and serves as a representative of the radiative transfer model.

**Pure ML model.** The pure ML model employs a deep neural network (DNN), where the optimal structure is filtered (as shown in Extended Data Fig. 1a) and the input features include $L_{sen}$, $\varepsilon$, and $\omega$ for LST retrieval. The network utilizes sigmoid activation functions in hidden layers to introduce non-linearity, complemented by initialization applied for the weight matrices to mitigate vanishing gradients. It is optimized by adaptive moment estimation (Adam) with a learning rate of 0.001. The model validation followed a leave-one-out cross-validation (LOO-CV) protocol, where training data from 26 stations ($n$ = 27 total sites) were utilized and the performance was evaluated on the excluded station.

**MM-ML coupled model.** The coupled model proposed in this paper combines the respective advantages of MM and ML models. The core of the LST retrieval model is to accurately capture the radiative physical processes between the atmosphere and the surface. In the simulation sample generation, we utilized the GAPRI and TIGR atmospheric profiles databases through the MODTRAN model to calculate the atmospheric spectral transmittance and radiance along the line-of-sight path, thereby generating a high-precision simulated dataset. MODTRAN simulates the atmospheric radiative transfer processes using a narrowband model, comprehensively accounting for absorption, emission, and scattering effects from molecules and particulates[50]. By integrating the spectral response function of Landsat 8, a link is established between the sensor radiance and atmospheric parameters to construct multi-source input data for the PCMCNN.

To overcome the limitations of the traditional empirical coefficient methods in atmospheric function parameter estimation, the DDRL was first designed in the PCMCNN, which includes three parallel DNN sub-networks that model the nonlinear relationships between the atmospheric function parameters $\psi_1$, $\psi_2$, and $\psi_3$ and the atmospheric water vapor content based on SC atmospheric processes. The PPGL is introduced to encode the explicit physical relationships defined in [Eq. 6](#) into the network architecture to improve the model accuracy. To enhance the physical consistency among the atmospheric function parameters and improve the model adaptability, the PPOL is introduced. This layer utilizes the output of the LST from the RTE([Eq. 8](#)) to construct an energy function that uniformly constrains and optimizes the coupling between the atmospheric function parameters and LST. Through the PPOL, the framework can effectively capture the complex atmosphere-surface interaction mechanisms while ensuring that the predictions are consistent with the physics of radiative transfer. The weights and biases of the DNN are optimized by a back-propagation algorithm to gradually reduce the prediction error while improving the generalization ability. To avoid overfitting, we systematically evaluated the performances of different network structures (as shown in [Extended Data Fig. 1b](#)), and ultimately determined the optimal configuration. The activation function was the sigmoid activation function.

The well-trained coupled model was applied to the global LST retrieval task. By inputting $L_{sen}$, ε, and $\omega$, the MM-ML model outputs LST, achieving an end-to-end retrieval process. By leveraging the nonlinear fitting capabilities of neural networks and the physical principles of MMs, the framework can accurately predict global LST.

**References:**

1. Tong, Y. et al. Global Lakes are Warming Slower than Surface Air Temperature Due to Accelerated Evaporation. *Nature Water* **1**, 929-940 (2023).
2. Li, Z. et al. Satellite Remote Sensing of Global Land Surface Temperature: Definition, Methods, Products, and Applications. *Rev. Geophys.* **61**, e2022RG000777 (2023).
3. Wang, A., Zhang, M., Chen, E., Zhang, C. & Han, Y. Impact of Seasonal Global Land Surface Temperature (LST) Change On Gross Primary Production (GPP) in the Early 21St Century. *Sust. Cities Soc.* **110**, 105572 (2024).
4. Massaro, E. et al. Spatially-Optimized Urban Greening for Reduction of Population Exposure to Land Surface Temperature Extremes. *Nat. Commun.* **14**, 2903 (2023).
5. Zhong, Z. et al. Reversed Asymmetric Warming of Sub-Diurnal Temperature Over Land During Recent Decades. *Nat. Commun.* **14**, 7189 (2023).
6. Yin, J. et al. Future Socio-Ecosystem Productivity Threatened by Compound Drought–Heatwave Events. *Nat. Sustain.* **6**, 259-272 (2023).
7. Li, J. & Thompson, D. W. J. Widespread Changes in Surface Temperature Persistence Under Climate Change. *Nature* **599**, 425-430 (2021).
8. Li, Y. et al. Observed Different Impacts of Potential Tree Restoration On Local Surface and Air Temperature. *Nat. Commun.* **16**, 2335 (2025).
9. Li, Z. L. et al. Satellite-Derived Land Surface Temperature: Current Status and Perspectives - ScienceDirect. *Remote Sens. Environ.* **131**, 14-37 (2013).
10. Townshend, J. R. G., Justice, C. O., Skole, D., Malingreau, J. P. & Ruttenberg, S. The 1 Km Resolution Global Data Set: Needs of the International Geosphere Biosphere Programme. *Int. J. Remote Sens.* **15**, 3417-3441 (1994).
11. Chen, D., Huazhong, R., Qiming, Q., Jinjie, M. & Shaohua, Z. A Practical Split-Window Algorithm for Estimating Land Surface Temperature From Landsat 8 Data. *Remote Sens.* **7**, 647-665 (2015).
12. Jiménez-Muñoz, J. C., Sobrino, J. A., Skokovic, D., Mattar, C. & Cristóbal, J. Land Surface Temperature Retrieval Methods From Landsat-8 Thermal Infrared Sensor Data. *IEEE Geosci. Remote Sens. Lett.* **11**, 1840-1843 (2014).
13. Sobrino, J. A. & Romaguera, M. Land Surface Temperature Retrieval From MSG1-SEVIRI Data. *Remote Sens. Environ.* **92**, 247-254 (2004).
14. Jiménez-Muñoz, J. C. & Sobrino, J. A. A Generalized Single-Channel Method for Retrieving Land Surface Temperature From Remote Sensing Data. *Journal of Geophysical Research Atmospheres* **108**, - (2003).
15. Qin, Z., Karnieli, A. & Berliner, P. A Mono-Window Algorithm for Retrieving Land Surface Temperature From Landsat TM Data and its Application to the Israel-Egypt Border Region. *Int. J. Remote Sens.* (2001).
16. Gillespie, A. R. A. E. Residual Errors in ASTER Temperature and Emissivity Standard Products AST08 and AST05. *Remote Sensing of Environment: An Interdisciplinary Journal* **115** (2011).
17. Wang et al. Temperature and Emissivity Retrievals From Hyperspectral Thermal Infrared Data Using Linear Spectral Emissivity Constraint. *IEEE Transactions On Geoscience & Remote Sensing* (2011).
18. Sòria, G. & Sobrino, J. A. ENVISAT/AATSR Derived Land Surface Temperature Over a Heterogeneous Region. *Remote Sens. Environ.* **111**, 409-422 (2007).

19. Jiang, G. M., Li, Z. L. & Nerry, F. O. Land Surface Emissivity Retrieval From Combined Mid-Infrared and Thermal Infrared Data of MSG-SEVIRI. *Remote Sens. Environ.* **105**, 326-340 (2006).
20. Sobrino, J. A., Jiménez-Muoz, J. C. & Paolini, L. Land Surface Temperature Retrieval From LANDSAT TM 5. *Remote Sens. Environ.* **90**, 434-440 (2004).
21. Shen, H. & Zhang, L. Mechanism-Learning Coupling Paradigms for Parameter Inversion and Simulation in Earth Surface Systems. *Science China Earth Sciences* **66**, 568-582 (2023).
22. Camps-Valls, G. et al. Artificial Intelligence for Modeling and Understanding Extreme Weather and Climate Events. *Nat. Commun.* **16**, 1919 (2025).
23. Kochkov, D. et al. Neural General Circulation Models for Weather and Climate. *Nature* **632**, 1060-1066 (2024).
24. Bi, K. et al. Accurate Medium-Range Global Weather Forecasting with 3D Neural Networks. *Nature* **619**, 533-538 (2023).
25. Bauer, P., Dueben, P. D., Hoefler, T., Quintino, T. & Wedi, N. P. The Digital Revolution of Earth-System Science. *Nature Computational Science* **1**, 104-113 (2021).
26. Markus et al. Deep Learning and Process Understanding for Data-Driven Earth System Science. *Nature* (2019).
27. Bergen, K. J., Johnson, P. A., de Hoop, M. V. & Beroza, G. C. Machine Learning for Data-Driven Discovery in Solid Earth Geoscience. *Science* **363**, eaau0323 (2019).
28. Ganguly, A. R. et al. Toward Enhanced Understanding and Projections of Climate Extremes Using Physics-Guided Data Mining Techniques. *Nonlinear Process Geophys.* **1**, 51-96 (2014).
29. Chantry, M., Christensen, H., Dueben, P. & Palmer, T. Opportunities and Challenges for Machine Learning in Weather and Climate Modelling: Hard, Medium and Soft AI. *Philosophical Transactions of the Royal Society a Mathematical Physical and Engineering Sciences* (2021).
30. Ma, J. et al. A Mechanism-Guided Machine Learning Method for Mapping Gapless Land Surface Temperature. *Remote Sens. Environ.* **303** (2024).
31. Plésiat, É., Dunn, R. J. H., Donat, M. G. & Kadow, C. Artificial Intelligence Reveals Past Climate Extremes by Reconstructing Historical Records. *Nat. Commun.* **15**, 9191 (2024).
32. Wang, F., Tian, D. & Lowe, L. L. J. Deep Learning for Daily Precipitation and Temperature Downscaling. *Water Resour. Res.* **57**, e2020W-e29308W (2021).
33. Noori, N., Kalin, L. & Isik, S. Water Quality Prediction Using SWAT-ANN Coupled Approach. *J. Hydrol.*, 125220 (2020).
34. Shen, H., Jiang, Y., Li, T., Cheng, Q. & Zhang, L. Deep Learning-Based Air Temperature Mapping by Fusing Remote Sensing, Station, Simulation and Socioeconomic Data. *Remote Sens. Environ.* **240**, 111692 (2020).
35. Mao, K. et al. A General Paradigm for Retrieving Soil Moisture and Surface Temperature From Passive Microwave Remote Sensing Data Based On Artificial Intelligence. *Remote Sens.* **15**, 1793 (2023).
36. Wang, X., Zhong, L. & Ma, Y. Estimation of 30 M Land Surface Temperatures Over the Entire Tibetan Plateau Based On Landsat-7 ETM+ Data and Machine Learning Methods. *Int. J. Digit. Earth* **15**, 1038-1055 (2022).
37. Wang, H. et al. A Method for Land Surface Temperature Retrieval Based On Model-Data-Knowledge-Driven and Deep Learning. *Remote Sens. Environ.* **265**, 112665 (2021).

38. Cheng, Y. et al. A Robust Framework for Accurate Land Surface Temperature Retrieval: Integrating Split-Window Into Knowledge-Guided Machine Learning Approach. *Remote Sens. Environ.* **318**, 114609 (2025).
39. Malakar, N. K. et al. An Operational Land Surface Temperature Product for Landsat Thermal Data: Methodology and Validation. *IEEE Transactions On Geoence and Remote Sensing*, 5717-5735 (2018).
40. Mattar, C. et al. Global Atmospheric Profiles From Reanalysis Information (GAPRI): A New Database for Earth Surface Temperature Retrieval. *Int. J. Remote Sens.* **36**, 5045-5060 (2015).
41. Chevallier, F., Chéruy, F., Scott, N. A. & Chédin, A. A Neural Network Approach for a Fast and Accurate Computation of a Longwave Radiative Budget. *Journal of Applied Meteorology* **37**, 1385-1397 (1998).
42. Chedin, A., Scott, N. A., Wahiche, C. & Moulinier, P. The Improved Initialization Inversion Method: A High Resolution Physical Method for Temperature Retrievals From Satellites of the TIROS-N Series. *Journal of Applied Meteorology* **24**, 128-143 (1985).
43. Driemel, A. et al. Baseline Surface Radiation Network (BSRN): Structure and Data Description (1992--2017). *Earth Syst. Sci. Data* **10**, 1491-1501 (2018).
44. Augustine, J. A., Deluisi, J. J. & Long, C. N. SURFRAD—a National Surface Radiation Budget Network for Atmospheric Research. *Bull. Amer. Meteorol. Soc.* **81**, 2341-2357 (2000).
45. Yu, W., Ma, M., Li, Z., Tan, J. & Wu, A. New Scheme for Validating Remote-Sensing Land Surface Temperature Products with Station Observations. *Remote Sens.* **9**, 1210 (2017).
46. Yu, W. et al. Evaluation of MODIS LST Products Using Longwave Radiation Ground Measurements in the Northern Arid Region of China. *Remote Sens.* **6**, 11494-11517 (2014).
47. Yu, W., Ma, M., Wang, X., Song, Y. & Tan, J. Validation of MODIS Land Surface Temperature Products Using Ground Measurements in the Heihe River Basin, China. *Proc. SPIE* **8174**, 772-783 (2011).
48. Wang, K. et al. Estimation of Surface Long Wave Radiation and Broadband Emissivity Using Moderate Resolution Imaging Spectroradiometer (MODIS) Land Surface Temperature/Emissivity Products. *Journal of Geophysical Research: Atmospheres* **110** (2005).
49. Cook, M., Schott, J. R., Mandel, J. & Raqueno, N. Development of an Operational Calibration Methodology for the Landsat Thermal Data Archive and Initial Testing of the Atmospheric Compensation Component of a Land Surface Temperature (LST) Product From the Archive. *Remote Sens.* **6**, 11244-11266 (2014).
50. Berk, A., Anderson, G. P., Acharya, P. K., Bernstein, L. S. & Lewis, P. E. MODTRAN5: A Reformulated Atmospheric Band Model with Auxiliary Species and Practical Multiple Scattering Options. *Proc. SPIE* **5425**, 341-347 (2004).

**Data availability**

All data used in this study are publicly available: TIGR atmospheric profile dataset: (https://ara.lmd.polytechnique.fr/index.php?page=tigr). GAPRI atmospheric profile dataset: (http://biosfera.uchile.cl/descargas/GAPRI-v1.6.zip). Landsat 8 data (https://earthexplorer.usgs.gov/). NECP reanalysis data: (https://psl.noaa.gov/data/gridded/data.ncep.reanalysis.html). ASTER GEDv3 data: (Index of /ASTT/AG100.003). Heihe Integrated Observatory Network site data: (https://data.tpdc.ac.cn/projectDataList?specialId=4e7b97a6-ee1f-4b51-83e5-04a100993973). SURF RAD site data: (http://gml.noaa.gov/grad/surfrad/). BSRN site data: (https://bsrn.awi.de/).

**Code availability**

Codes to reproduce the study are available via GitHub at
https://github.com/TianXie-WHU/PCMCNN.git

**Acknowledgements**

This work was supported by the Key Program of the National Natural Science Foundation of China under grant no. 42130108. The numerical calculations in this paper were performed on the supercomputing system at the Supercomputing Center of Wuhan University.

**Author contributions**

All authors made contributions to this paper. H. S. and L. Z. initiated and supervised the project. T.X. and M.J. conceived the research and conducted the experiments. H.L., C.Z., J.M., and G.Z. were involved in the discussion. H.L., C.Z., and X.G. oversaw the research progress. All authors edited the manuscript.

**Competing interests**

The authors declare no competing interests.

a

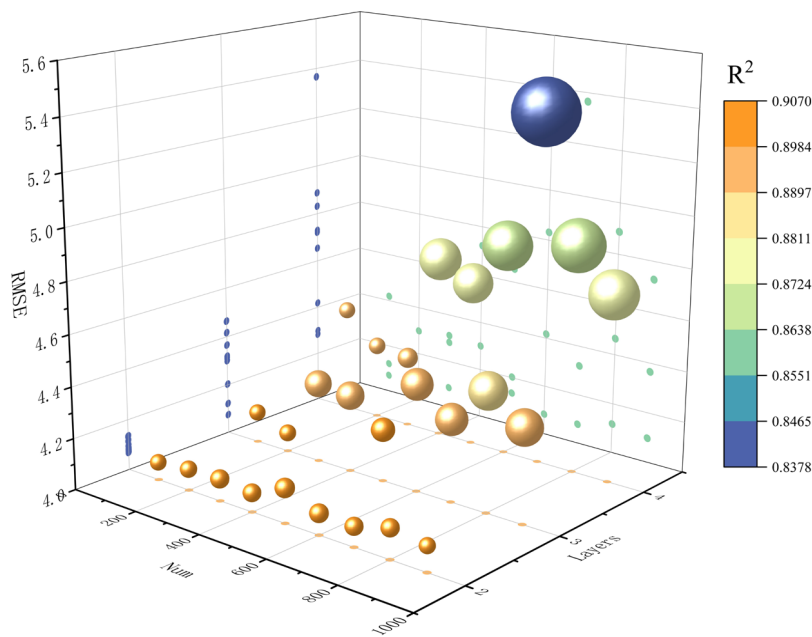

b

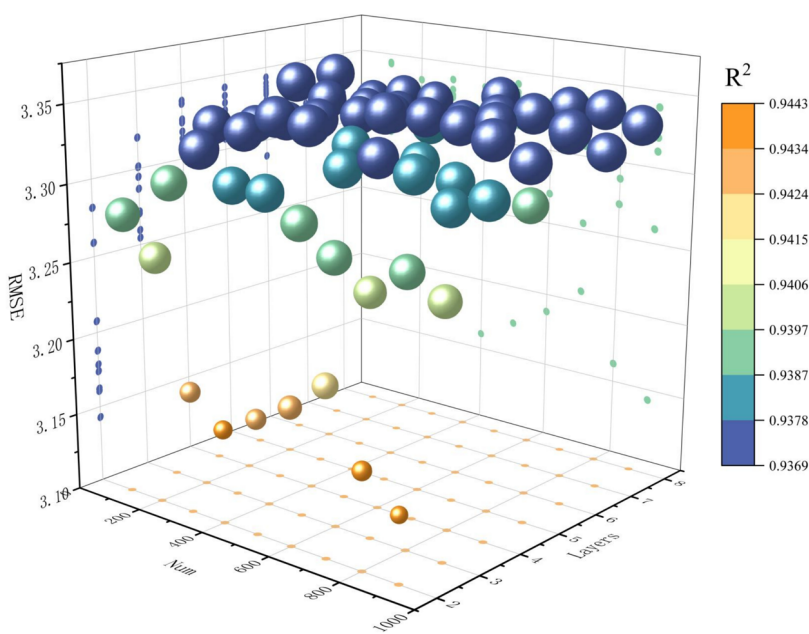

**Extended Data Fig. 1 | Site validation results for the pure ML and the MM-ML models with different network configurations.** Different combinations of hidden layers (from 2 to 8 with an interval of 1) and the number of neurons (from 100 to 900 with an interval of 100) in the deep neural network were analyzed. The icon size represents the MAE value, the Z-axis indicates the RMSE value, and the color reflects the R-value. **a,** Scatterplot of the pure ML model based on the site leave-one-out cross-validation accuracy at the global site data level, as the number of deep neural network layers and the number of neurons varies. **b,** Scatterplot of the MM-ML model based on the site-independent validation accuracy at the global site dataset, varying with the number of deep neural network layers and the number of neurons. In both screenings, we found that neural networks with a shallower number of layers were able to obtain higher accuracy.

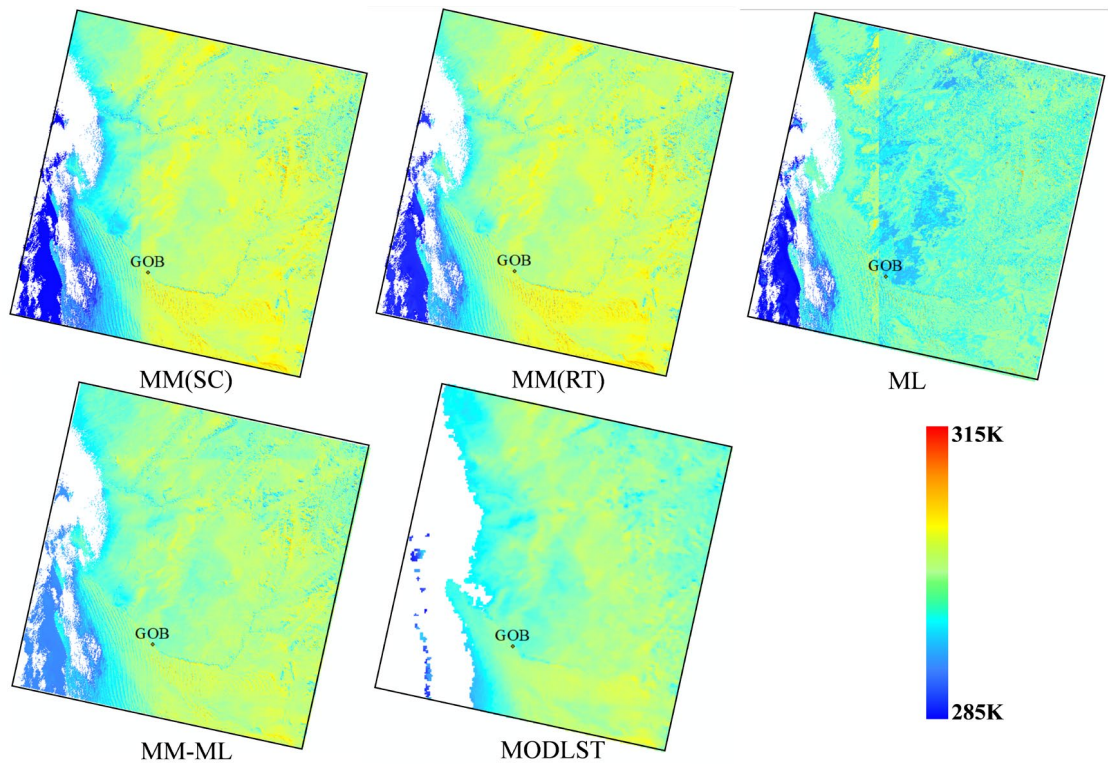

**Extended Data Fig. 2 | Validation of the MM(SC), MM(RT), pure ML, and MM-ML models under low atmospheric water vapor content.** Validation results of the four models (MM(SC), MM(RT), pure ML, and MM-ML) compared with the MODIS LST product (MODLST) at the GOB site in Africa on June 17, 2023, under low atmospheric water vapor content (0.403 g/cm²).

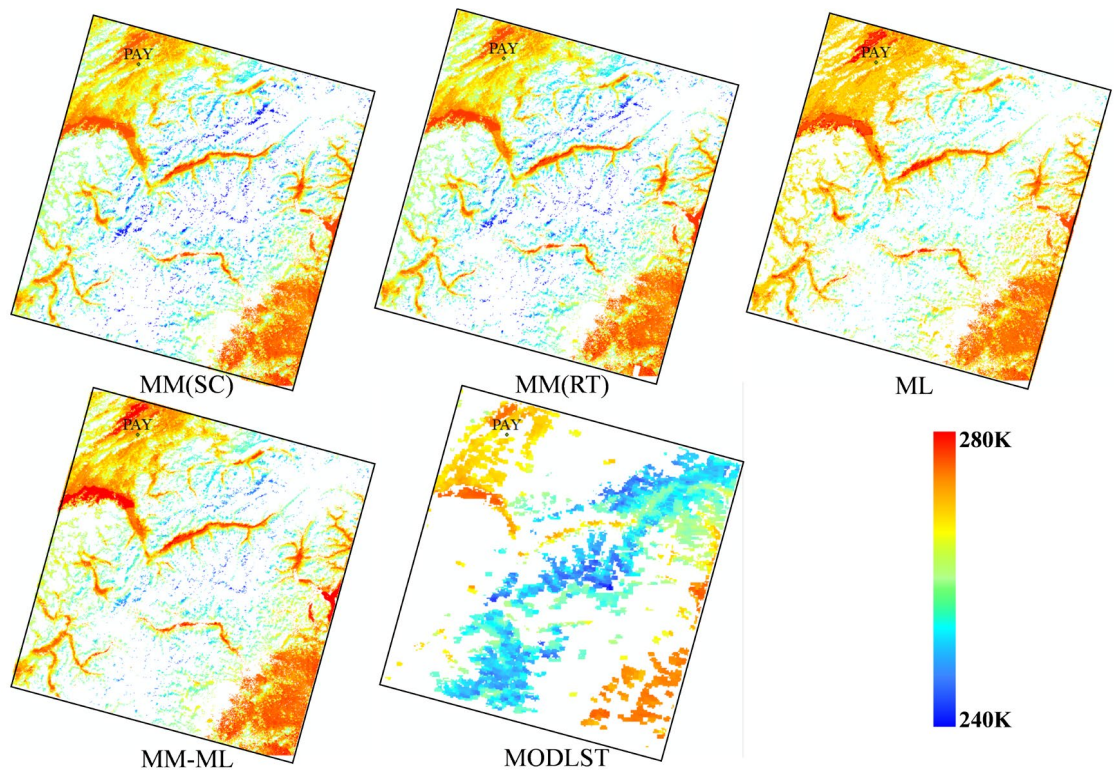

**Extended Data Fig. 3 | Validation of the MM(SC), MM(RT), pure ML, and MM-ML models under low surface temperature conditions.** Validation results of the four models (MM(SC), MM(RT), pure ML, and MM-ML) compared with the MODIS LST product (MODLST) at the PAY site in Europe on February 27, 2018, under low surface temperature conditions (268.968 K).

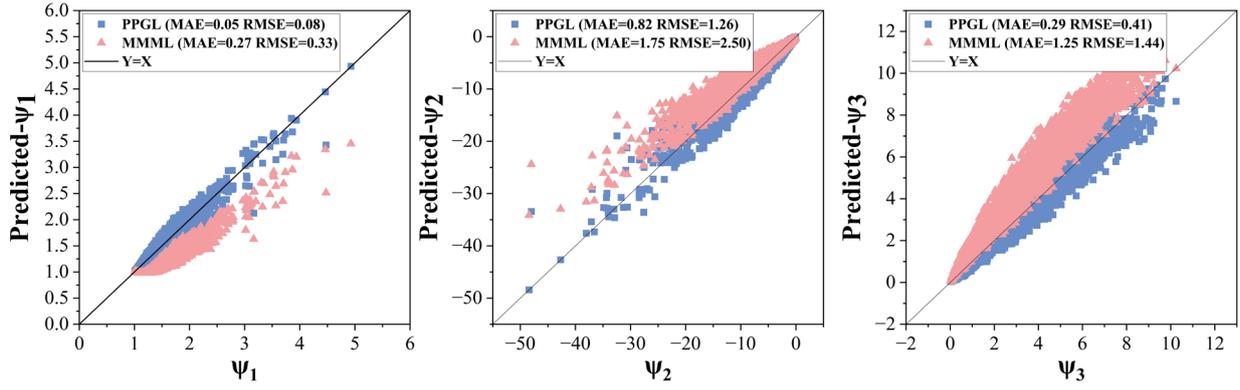

**Extended Data Fig. 4 | Scatterplots of the predicted atmospheric function parameters from the single PPGL and the MM-ML model, compared with the label data.** This figure shows the results of predicting the three atmospheric function parameters using the single PPGL and the MM-ML model. The scatterplots compare the predictions with the label data.

**Extended Data Table 1 | Basic information for the global sites.**

| Area | Code | Name | Location | Num |
|---|---|---|---|---|
| Asia | AR | A'rou | 38.0473N, 100.4643E | 66 |
| | CAP | Cape Baranova | 79.27N, 101.75E | 12 |
| | DM | Daman | 38.8555N, 100.3722E | 83 |
| | HMZ | Desert | 42.1135N, 100.9872E | 165 |
| | HZZ | Huazhaizi Desert steppe | 38.7659N, 100.3201E | 92 |
| | SDQ | Sidaoqiao | 42.0012N, 101.1374E | 167 |
| | TAT | Tateno | 36.0581N, 140.1258E | 78 |
| | YK | Yakou | 38.0142N, 100.2421E | 24 |
| | ZYSD | Zhangye wetland | 38.9751N, 100.4464E | 81 |
| Europe | BUD | Budapest-Lorinc | 47.4291N, 19.1822E | 59 |
| | CAB | Cabauw | 51.968N, 4.928E | 116 |
| | NYA | Ny-Ålesund | 78.9227N, 11.9273E | 141 |
| | PAY | Payerne | 46.8123N, 6.9422E | 247 |
| | TOR | Toravere | 58.2641N, 26.4613E | 65 |
| North America | ALE | Alert | 82.49N, 62.42W | 3 |
| | BAR | Barrow | 71.323N, 156.607W | 43 |
| | BND | Bondville | 40.0519N, 88.3731W | 168 |
| | DRA | Desert Rock | 36.6237N, 116.0195W | 273 |
| | FPK | Fort Peck | 48.3078N, 105.1017W | 277 |
| | GWN | Goodwin Creek | 34.2547N, 89.8729W | 200 |
| | PSU | Penn. State Univ | 40.7201N, 77.9308W | 73 |
| | SEL | Selegua, Mexico Solarimetric | 15.784N, 91.9902W | 20 |
| | SXF | Sioux Falls | 43.7340N, 96.6233W | 208 |
| | TBL | Table Mountain, Boulder | 40.1250N, 105.2368W | 230 |
| Africa | GOB | Gobabeb | 23.5614S, 15.042E | 192 |
| | IZA | Izaña | 28.3093N, 16.4993W | 95 |
| South America | OHY | Observatory of Huancayo | 12.05S, 75.32W | 7 |

**Extended Data Table 2 | Single PPGL and MM-ML model simulation dataset and site dataset validation accuracy.**

| Validation method | Method | MAE | RMSE | $R^2$ |
|---|---|---|---|---|
| Simulation datasets | PPGL | 1.538 | 2.363 | 0.983 |
| | MM-ML | **0.721** | **1.121** | **0.996** |
| Global site datasets | PPGL | 3.299 | 4.109 | 0.905 |
| | MM-ML | **2.375** | **3.061** | **0.947** |